\documentclass[a4paper]{article}

\usepackage{INTERSPEECH2020}
\usepackage{xcolor,graphicx,amssymb,multirow,multicol,booktabs}
\usepackage{siunitx}
\usepackage{todonotes}
\usepackage{tikz}
\usetikzlibrary{shapes,arrows}

\title{Large scale weakly and semi-supervised learning for low-resource video ASR}
\name{\begin{tabular}{c}Kritika Singh*, Vimal Manohar*, Alex Xiao*, Sergey Edunov, Ross Girshick, Vitaliy Liptchinsky,\\Christian Fuegen, Yatharth Saraf, Geoffrey Zweig, Abdelrahman Mohamed\thanks{*Equal contribution}\end{tabular}}
\address{Facebook AI}
\email{\{skritika,vimalmanohar,axiao,edunov,rbg,vitaliy888,fuegen,ysaraf,gzweig,abdo\}@fb.com}

\begin{document}
\maketitle
\begin{abstract}
Many semi- and weakly-supervised approaches have been investigated for overcoming the labeling cost of building high-quality speech recognition systems. On the challenging task of transcribing social media videos in low-resource conditions, we conduct a large scale systematic comparison between two self-labeling methods on one hand, and weakly-supervised pre-training using contextual metadata on the other. We investigate distillation methods at the frame level and the sequence level for hybrid, encoder-only Connectionist Temporal Classification (CTC) based, and encoder-decoder speech recognition systems on Dutch and Romanian languages using 27,000 and 58,000 hours of unlabeled audio respectively. Although all approaches improved upon their respective baseline word error rates (WER) by more than 8\%, sequence-level distillation for encoder-decoder models provided the largest relative WER reduction of 20\% compared to the strongest data-augmented supervised baseline.
\end{abstract}

\section{Introduction}
Recent advances in speech recognition systems have enabled successful large scale deployments of various customer-facing speech applications, e.g. personal conversational agents, automatic transcriptions for accessibility, and multi-modal video understanding. This trend has increased the need for developing accurate automatic speech recognition (ASR) models for many languages and domains. However, there are significant challenges to achieving this. 

First, even though current ASR systems are arguably within striking distance of human performance for broadcast news and telephone-speech domains \cite{amodei2016deep, saon2017english, xiong2017achieving}, more challenging real-world scenarios involving unconstrained, natural speech that is filled with background music and noise, various speaking styles and emotions, disfluencies, heavy accents, un-cued speaker and language switching, is still an open problem for speech recognition systems \cite{ibm_malach_19}. In this paper, we focus on the domain of public social media videos which involve all these challenges while representing an interesting benchmark for evaluating the effectiveness of different learning methods due to their ever increasing amount, multi-modal nature, and the availability of related metadata, e.g. video title, post text, and comments.

Second, it can be prohibitively expensive and difficult to collect sufficient amounts of supervised training labels to feed data hungry neural speech recognition models for each language. Therefore, we'd like to minimize the amount of supervised data used in training these models. With increased access to large computational resources, three families of methods have emerged in this direction: (1) Using large volumes of unpaired audio and text data. \cite{wessel_05, hori2019cycle, cpc, effectiveness2019} (2) Augmenting audio data with contextual metadata as distant labels \cite{csl01_limsi, glass_16, ttic_may2017, singh2019Spuru}. (3) Data augmentation through reverberation, structured noise, speed perturbation, time and frequency masking \cite{kaldi_augment, spec_augment}. Given the various existing approaches for combining unlabeled audio with unpaired text and for incorporating distant labels, it is unclear, however, which of these methods are complementary, how they compare to each other, and how they can be combined effectively into a scalable recipe to maximize speech recognition performance for a complex domain like social media videos under low-resource constraints. 

In this paper, we take a step towards answering some of these questions by conducting large scale experiments to compare and combine methods drawn from these families of techniques, focusing on low-resource setups for transcribing public social media videos in two languages: Dutch and Romanian, using 27,000 and 58,000 hours of unlabeled data respectively. Applying data augmentation for all models \cite{kaldi_augment, spec_augment}, we compare a recently proposed weakly-supervised approach \cite{singh2019Spuru} (Section \ref{weak-supervision}) and two commonly used self-labeling methods: frame-level distillation \cite{hari_2019} and sequence-level distillation \cite{wessel_05, tara_sslearning} for hybrid, CTC-based, and encoder-decoder ASR setups (Section \ref{self-labeling}).

\section{Weakly-supervised Speech Recognition} \label{weak-supervision}
% moved from intro
% Attention-based encoder-decoder models were shown to be able to harness the loose relationship and alignment between the video metadata, e.g. title and comments, and the audio signal with varying degrees of success depending on the relevance of the metadata to the spoken content \cite{singh2019Spuru}. One reason for the observed gains is the inclusion of much more acoustic diversity than those present in the limited labeled data, even with such noisy distant supervision.\\
Following \cite{singh2019Spuru}, we use video metadata, e.g. title and post text, as distant labels for the ASR task. We use two sets of training data: (i) $\{X, Y^s\} \in \mathcal{D}^s$ is the supervised set where $X$ and $Y^s$ are pairs of audio features and label sequences. (ii) $\{X, Y^w\} \in \mathcal{D}^w$ is the weakly-supervised set where $X$ and $Y^w$ are pairs of audio features and the corresponding contextual text. The targets $Y^s$ and $Y^w$ are sequences of sub-word units \cite{sp}.

An encoder-decoder approach \cite{las, s2s_speech_MO} is used for maximizing the conditional probability of generating the contextual text sequence $Y^w$ given $X$, an input sequence of mel-scale log filterbank features, where $x_i \in R^d$
\begin{equation}
\begin{split}
    \mathcal{F}^w &= p(Y^w|X; \theta^w) \\
    &=\prod_{i=1}^M p(y_i^w | y_1^w, y_2^w, ..., y_{i-1}^w, x_1, x_2, ..., x_T; \theta^w)
\end{split}
\end{equation}
The attention-based encoder-decoder framework fits well with the weak supervision since it offers flexible alignment and unconstrained coverage between input and output sequences. Other ASR training approaches aren't as suitable given the abstractive relationship between $Y^w$ and $X$. Hybrid Hidden Markov Model-Neural Network (HMM-NN) and CTC approaches assume either a frame-level or a monotonic alignment between input and target sequences, and they constrain the possible length of the target sequence by the input sequence length.  
%The hybrid HMM-NN approach requires a low-level sub-second alignment between input audio and output targets, while the CTC approach assumes a monotonic alignment between input and target sequences, and it constrains the maximum possible length of the target sequence by the length of the input sequence.

The final objective function is $\mathcal{F} = \mathcal{F}^w + \mathcal{F}^s$ where the supervised term, $\mathcal{F}^s = p(Y^s|X; \theta^s)$, is also maximized using the encoder-decoder approach. We share the full model for both types of data, where $\theta^w = \theta^s = \{\theta_{enc}, \theta_{dec}\}$ combines the parameters in the audio encoder and the language generation and attention modules in the decoder. During training, we alternate, with some mixing ratio, between mini-batches sampled from the two training sets $\mathcal{D}^s$ and $\mathcal{D}^w$.

\section{Self-labeled Speech Recognition} \label{self-labeling}
% Semi-supervised learning augments the small amount of labeled data with a large amount of unlabeled data to improve the overall system performance, for example, by utilizing the neighborhood structure of data points\cite{}, or through learning a conditional generative model of input audio \cite{}. 
Self-labeling is one of the most effective methods of semi-supervised learning for speech recognition \cite{Kemp1999, csl01_limsi, wessel_05}, where a teacher model with limited supervision extends the training data by labeling an extra amount of unlabeled data.
% supervised teacher model trained with a limited transcribed data is used for labeling the extra unlabeled data and include them in the standard supervised learning procedure. 
% To improve the quality of self-labels, multiple teachers are often combined during learning and confidence score based filtering \cite{ma_bbn_06} is used to remove ambiguous segments. 
In this study, we investigate two approaches:

\subsection{Frame-level distillation}
%Early variants of this approach date back to the 70s \cite{}, and to GMM-HMM speech recognition systems where frame-level alignments, i.e. the top frame-level hypothesis, from one model are used as labels during another iteration of model training. 

% The hybrid ASR modeling paradigm leverages an acoustic model that maps audio input features to a distribution over phonetic or graphemic output units. These are typically trained on force-aligned targets from ground truth human transcriptions. \cite{hinton_15} proposed training acoustic models with knowledge distillation: optimizing a student model's frame-level probability distribution to be the same as a teacher model's. Since knowledge distillation doesn't require labels, it can be applied over a large volume of unlabeled data \cite{hari_2019}. The objective is the KL divergence of the frame-level distributions of the teacher and student, which is equivalent to minimizing:
In hybrid ASR modeling, acoustic models are trained to map input audio features into phonetic or graphemic output units by minimizing the frame-level cross entropy between the model predictions and ground truth labels. %labels given by a forced alignment with the ground truth transcription. 
Frame-level knowledge distillation \cite{caruana_06,hinton_15} replaces the ground truth labels with probability distributions generated from a teacher model over all output units for the same set of supervised data, or for a large volume of unlabeled audio \cite{hari_2019}. It minimizes the KL divergence between outputs of the model, referred to as a student, and the teacher model's frame-level distributions:
\begin{equation}
\vspace{-0.1cm}
% \begin{split}
    \mathcal{F}^d = -\frac{1}{T} \sum_{t=1}^T \sum_{k=1}^K p(y_{t}=k|X) \log \hat{p}(y_{t}=k|X), 
%\vspace{-0.1cm}
% \end{split}
\end{equation}
where $X$ is the sequence of features with $T$ time-steps, $K$ is the number of classes, $p(y_{t} = k|X)$ is the teacher frame-level probability for class $k$ and time step $t$, and $\hat{p}(y_{t} = k|X)$ is the student frame-level probability. 

%The formulation of this approach is commonly applied to hybrid HMM-NN systems which generate frame-level posterior distributions, and forces both the student and teacher models to have identical output units. 
% To address early shortcoming of not using the language model information during teacher-student learning, lattice-based self-labeling techniques are investigated \cite{vimal_18}. Alex: commented this out to save space because we also use LM info in sequence level distillation and we cited this paper elswhere already

\subsection{Sequence-level distillation} 
% Given 1-best hypotheses generated by a teacher model as target sequences, a student model is trained using a large volume of unlabeled audio data. One potential advantage of this technique over the frame-level distillation approach is the incorporation of a language model (LM) when generating the 1-best hypotheses. This LM can be adapted to a specific genre or domain during decoding and can also be used to filter out audio segments with very unlikely word sequences. 
In sequence-level distillation, the student model attempts to learn the teacher model's distribution over the entire sequence as opposed to just the individual frames. The full teacher sequence distribution is often intractable, but it can be approximated with a lattice \cite{sequence_ts} for hybrid or the top k hypotheses for CTC \cite{ctc_ts} and seq2seq \cite{seq2seq_ts, seq2seq_ts_spec_aug}. 
Here, we consider an approximation by just training on the top-1 hypotheses generated by the teacher model.  
Unlike frame-level distillation, sequence-level distillation utilizes a language model (LM)
%\footnote{The language model can be implicit in seq2seq model, or can be an explicitly provided external model} 
during label generation. The LM can be adapted to a specific genre or domain, and can also be used to filter out audio segments with very unlikely word sequences. Sequence-level distillation is the first form of self-labelling applied for speech recognition \cite{Kemp1999, csl01_limsi, ma_bbn_06, wessel_05}. It's also commonly used for non-autoregressive machine translation systems \cite{gu2017non}.
%\secomment{In MT this technique is known as "sequence-level knowledge distillation", see paragraph 3.2 here https://arxiv.org/pdf/1606.07947.pdf they also argue that in autoregressive models this technique allows to better transfer knowledge on the sequence level compared to regular KD }

When using target word sequences instead of frame-level probability distributions, the student system has greater flexibility to (a) Follow any modelling paradigm, e.g. hybrid, encoder-only with CTC loss, or attention/transducer-based encode-decoder % attention/transducer-based seq2seq, 
(b) Use an output vocabulary for the student that is different from the teacher ASR system, (c) Filter the unlabeled data and focus on specific weaknesses in the originally collected labeled data, e.g. improving recognition of named entities in a specific domain \cite{tara_sslearning} and, (d) Evaluate how synergistic self-labels and weak-supervision in the form of metadata labels are through multi-task learning of the encoder-decoder model with mini-batches sampled from both data sources which we will show in the experimental section.

\section{Experiments}
\label{experiments}
\subsection{Data}
\label{data}
We explore the aforementioned weakly-supervised and self-labeling approaches on two languages under low-resource conditions: Dutch and Romanian. Both languages have three test sets -- test-clean, test-noisy and test-extreme -- listed in increasing order of difficulty in acoustic conditions, and a single dev-noisy set. The sizes of the supervised sets -- train, dev-noisy, test-clean, test-noisy and test-extreme -- are 290, 13, 10, 25, and 12 hours respectively for Dutch. They are 163, 6, 6, 12, and 10 hours respectively for Romanian. All our datasets are sampled from public videos that are anonymized. The unlabeled audio data for both languages come from videos that are up to 5 minutes in duration. First, the whole data is decoded with a baseline hybrid model and segmented using hypothesized word boundaries into chunks at most 10s in duration while removing non-speech portions. \\
%for segmentation (max. of 10s) and removal of non-speech segments. 
This process yields the \texttt{data-large} set, which is about 27,000 hours for Dutch and 58,000 hours for Romanian. For our meta-data supervision experiments, these sets are further filtered to preserve videos with meta-data between 50 and 700 characters. Videos with meta-data that are completely orthogonal to the baseline ASR hypotheses (zero word overlap) are discarded. This process produces our weak-supervision data set \texttt{data-ws} of 7,900 hours for Dutch and 6,400 hours for Romanian. Note that the full metadata text is shared for different segments of the same video during weakly-supervised learning.

\subsection{Experimental setup}
% Restructure into:
%
% (a) Data Pre-processing
% feature extraction, data augmentation, segmentation, etc
%
% (b) System Description
%  acoustic model architecture
%
% (c) Training Details
% 
% CTC, encoder decoder, hybrid, hybrid distribution matching
%

% \begin{table}[]
%     \centering
   
%     \begin{tabular}{l|l|l|l}
%      \toprule
%         System & Hybrid & CTC & Enc-dec \\
%         \hline
%         Encoder & \multicolumn{3}{c}{LC-BLSTM} \\
%         \hline 
%         \\[-1em]
%         Decoder & \shortstack{5-gram\\ WFST} & \shortstack{5-gram \\ WFST} & \shortstack{VGG-\\Transformer} \\
%         \hline
%         Input features & \multicolumn{3}{c}{80-dim log-mel} \\
%         \hline
%         Window size & 25ms & 16ms & 16ms \\
%         \hline
%         Window shift & \multicolumn{3}{c}{10ms} \\
%         \hline
%         Output units & Chenone & 1000 wp & 1000 wp \\
%     \bottomrule
%     \end{tabular}
%     \caption{Experimental setup for the baseline systems}
%     \label{tab:my_label}
%     \vspace{-6mm}
% \end{table}

\subsubsection{Data Preprocessing}
% The input speech is encoded into 80 dimensions of mel-scale log filterbank features computed over a window of size \textit{w} and shift of 10ms. \textit{w} is 25ms for hybrid systems and 16ms for CTC and encoder-decoder systems.\\
% We augment our training data by applying speed perturbation and additive noise. For both the labeled and unlabeled sets, two copies of the original audio are generated by perturbing the speed by 0.9 and 1.1. These new copies combined with the unperturbed copy are then superimposed with 10s clips of diverse noise events to give six copies in total. We use both speed perturbation and additive noise in all the training stages for encoder-decoder and CTC models. For hybrid models, we use speed perturbation and additive noise in all the stages except during self-labeling. \footnote{We found limited additional benefit from them at this stage.} \\
% In all the training stages, we apply on-the-fly, time and frequency masking of spectral features \cite{spec_augment}. We use the SM policy from \cite{spec_augment}: 2 frequency masks of width 15, 2 time masks of width 70, and limit the time mask to be no more than 0.2 times the number of time steps.  Both these were found to be very beneficial when the amount of supervised data is small.
We use 80 dimensional mel-scale log filterbank features computed every 10ms over 25ms windows for hybrid models and 16ms for all other models. Both labeled and unlabeled data are augmented with copies using 0.9 and 1.1 speed perturbation \cite{kaldi_augment}, and then one more copy by superposition with 10s clips of diverse noise events. In all training stages, we apply on-the-fly time and frequency masking following the Switchboard Mild (SM) policy from \cite{spec_augment}\footnote{Our self-labeled hybrid models don't use speed and noise perturbation as initial experiments found them of limited benefit given the unlabeled data size.}.
\subsubsection{System Description}
% For comparison of training paradigms viz. hybrid, CTC and encoder-decoder, we keep the architecture of the encoder the same for all the systems -- a 5 layer, 800 hidden node latency controlled bi-directional LSTM (LC-BLSTM) neural network sub-sampled in the time-dimension by 2 after the first layer. The hybrid systems use chenone output units \cite{le2019senones}, with 8800 chenones in the case of Dutch and 9272 in the case of Romanian. The decision tree for the mapping from graphemes in context to the chenone output units used in the hybrid systems is obtained from the HMM-GMM flat-start training as described in Section \ref{sssec:training-details}. The CTC and encoder-decoder systems use 1000 sub-word output vocabulary \cite{sp}. 

% For decoding, both the hybrid and CTC systems use a WFST decoder with a 5-gram language model trained on the supervised transcripts. The encoder-decoder system has the decoder module as the implicit decoder. The decoder consists of 4 1-D convolutional layers with kernel size of 3 and 256 output features followed by 2 transformer blocks with 1k hidden dimensions, 16 self-attention heads and a 4k projection layer before the ReLU nonlinearity.
% At inference time, a simple beam search with a beam size of 20 is used for decoding to get the 1-best hypothesis.
To facilitate comparisons, all our hybrid, CTC-based, and encoder-decoder models have the same encoder architecture of 5 layer, 800 hidden units latency controlled bidirectional LSTM (LC-BLSTM) with a factor of 2 sub-sampling over time after the first layer. While an architecture search and/or using a different neural model, e.g. Transformers \cite{speech_transformer,conv_trans,trans_vs_rnn, wang2020}, may lead to better absolute WER numbers, we believe that the conclusions reached in our experiments are invariant to such modeling decisions.\\
The hybrid systems use 8800 output chenone units \cite{le2019senones} for Dutch and 9272 for Romanian models. The CTC-based and encoder-decoder models use 1000 sub-word output units \cite{sp} trained on the supervised transcripts.\\ 
For decoding, both the hybrid and CTC-based models use a Weighted Finite State Transducer (WFST) decoder with a 5-gram LM trained on the supervised transcripts. Following the architecture in \cite{singh2019Spuru}, the decoder module of the encoder-decoder models consists of 4 1-D convolutional layers with kernel size of 3 and 256 output features followed by 2 transformer blocks with 1k hidden dimensions, 16 self-attention heads and a 4k projection layer before the ReLU nonlinearity.
At inference time, a beam search with a beam size of 20 is used, with no language model fusion, to get the 1-best hypothesis.\\

%\begin{figure}
%    \centering
    %\includegraphics{\input{LaTeX/flowchart}}
    %\caption{Flowchart}
%    \label{fig:training_flow_chart}
%\end{figure}

\subsubsection{Training Details}
\label{sssec:training-details}

The encoder-decoder based self-labeling and weakly-supervised training experiments consist of three phases as described in \cite{singh2019Spuru}: (1) An initial supervised \texttt{burn-in} phase (2) A \texttt{train-main} phase that uses a mixture of supervised and the unlabeled data with either meta-data or self-labels as targets. (3) A final supervised-only \texttt{fine-tune} phase. During \texttt{train-main} we save checkpoints every 5k model updates and average the last 20 checkpoints to initialize the \texttt{fine-tune} phase. We use the same process for self-labeling using CTC. For both encoder-decoder and CTC, we use Adam \cite{adam} optimizer with a fixed learning rate of \num{4e-4}during \texttt{burn-in} and \texttt{train-main} phases, and \num{4e-5} during encoder-decoder \texttt{fine-tune} and \num{5e-5} during CTC \texttt{fine-tune} phase. Gradient norm clipping at 10.0 is used, where total gradients are scaled by the number of utterances in each mini-batch.\\
Our hybrid acoustic models are first trained with frame-level cross entropy (CE) then with the Lattice-free Maximum Mutual Information (LFMMI) \cite{lfmmi} criterion where frame-level supervision and numerator lattices are generated using a smaller bootstrapping hybrid system. We use a dropout of 0.5 and Adam optimizer with learning rates of 0.005 and 0.00001 for the CE and LFMMI stages with Incremental block distributed data parallelism (BMUF) \cite{bmuf}. The learning rate is halved when the held-out set doesn't improve at the end of each epoch. The supervised baseline is used for segmenting the unlabeled data and generating the self-labels for sequence-distillation experiments. \\
We chose the size of the teacher model for frame-level distillation experiments based on performance: 6 layers of LC-BLSTMs with 1000 hidden units for Dutch and 5 layers of 800 hidden units for Romanian. Following \cite{hari_2019}, we approximate the full teacher distribution with the top 3 predictions per frame, to minimize storage and network costs, which on average was enough to cover 99\% of the per-frame probability mass. After distillation, hybrid models are then fine-tuned with LFMMI on supervised data.

\subsection{Results}
% \subsubsection{Self-labeling}
% Table \ref{tab1} shows the effect of the self-labeling approach, either frame-level distillation or sequence distillation, on the final performance of self-labeled ASR systems. Within the sequence distillation approach, we show results for hybrid, CTC, and encoder-decoder systems. The first section in the table presents the best baseline systems for hybrid, CTC, and encoder-decoder models trained exclusively on the available supervised data with no pre-training but with the same data augmentation strategy used for the self-labeled systems. It is clear that all self-labeling recipes improve compared to their respective baselines, with relative WER improvements ranging from 7\% to 16\% for hybrid models and 30\% to 40\% for encoder-decoder models. We also observe several interesting trends from Table \ref{tab1}:
Table \ref{tab1} compares the two strategies for self-labeling on \texttt{data-large}, frame-level distillation for hybrid models and sequence-level distillation for all of hybrid, CTC-based and encoder-decoder models, with respect to their supervised baselines. Our strongest baselines for both Dutch and Romanian are the LFMMI fine-tuned hybrid models. For sequence-level distillations, all self-labeled systems provided relative improvements over their respective baselines with more than 12\% for hybrid, 20\% for CTC-based, and 37\% for encoder-decoder. The relative improvements on the Romanian language are on average 8\% higher than Dutch, which has about double the supervised data size, across all approaches. 

\begin{table}
\centering
\setlength\tabcolsep{2.0pt}
 \begin{tabular}{cccccccc} 
\toprule
 \multirow{2}{*}{} &\multicolumn{3}{c}{Dutch}&& \multicolumn{3}{c}{Romanian} \\ 
 \cline{2-4}\cline{6-8}
 & clean & noisy & extreme && clean & noisy & extreme \\
 \midrule
 \multicolumn{8}{c}{Supervised baseline}\\
 \midrule
 Hybrid (LFMMI) & 23.6 & 23.3 & 32.8 && 17.5 & 19.5 & 32.9 \\ 
 CTC & 26.7 & 26.3 & 36.8 && 20.8 & 22.2 & 38.3 \\ 
 Enc-Dec & 27.2 & 27.0 & 39.0 && 25.5 & 26.8 & 46.0 \\ 
 \midrule
%  \midrule
\multicolumn{8}{c}{Self-labeling using frame-level distillation}
\\
\midrule
 Top 3 (CE) & 23.3 & 22.8 & 32.3 && 15.7 & 17.9 & 29.6 \\ 
 + LFMMI & 21.3 & 20.7 & 29.8 && 14.7 & 17.0 & 28.8 \\ 
 Top 1 (CE) & 22.9 & 22.6 & 32.0 && 15.9 & 17.8 & 29.7 \\ 
  + LFMMI & 21.7 & 21.6 & 30.9 && 14.7 & 16.9 & 28.6 \\ 
%  \midrule
\midrule
\multicolumn{8}{c}{Self-labeling using sequence-level distillation} \\
\midrule
 Hybrid (CE) & 23.6 & 23.3 & 32.6 && 16.3 & 18.4 & 30.7 \\ 
 + LFMMI & 20.9 & 20.8 & 29.7 && 14.7 & 16.8 & 28.3 \\ 
 SL-LFMMI  & 22.1 & 21.8 & 31.4 && 15.6 & 17.6 & 29.5 \\ 
 CTC & 22.5 & 22.2 & 31.4 && 14.9 & 17.2 & 29.4 \\ 
 Enc-Dec & \textbf{18.4} & \textbf{18.5} & \textbf{27.9} && \textbf{13.1} & \textbf{15.6} & \textbf{27.3} \\ 
\bottomrule
\end{tabular}
\caption{WERs of different self-labeling approaches and ASR models on \texttt{data-large} relative to baseline supervised models on the three test sets for Dutch and Romanian. All + LFMMI models were seeded with the CE model from the previous row and use supervised data only except for the LFMMI row prefixed with SL- that uses self-labeling data.}
\label{tab1}
\vspace{-3mm}
\end{table}

The encoder-decoder model has two advantages over the other two systems. First, by relying entirely on sub-word units, it is an open vocabulary system with the ability to generate out-of-vocabulary words. Moreover, it learns a longer range language model on these sub-word units over the self-labels while observing the entire encoder output. Therefore, it provides 20\% on average compared the best baseline supervised system.

%Relative WER reductions are the largest for cleaner test set, and reduces towards more noisy ones for both languages which points to the impact of label quality on the expected final gains.

Within sequence-level distillation systems, self-labeled CTC-trained models are better than the self-labeled CE-trained hybrid model by 5\% on average but with LFMMI discriminative training the hybrid system leaps forward by about 4\%. Adding discriminative training techniques like sMBR \cite{smbr_povey} on top of the CTC-trained model would bring further gains which we leave for a future investigation. Confirming its sensitivity to transcription quality \cite{vimal_18, hari_2019}, using self-labels for LFMMI training, even with the large amount of audio data, is not as useful as using the much smaller supervised data in the discriminative training stage of hybrid models.

Interestingly, using Top-3 frame-level labels for distillation provided the same level of gains as the hybrid sequence-level distillation which utilizes a language model during label generation.  When we use only the Top-1, the performance stays almost the same for Romanian but degrades slightly after LFMMI fine-tuning in Dutch. %The Top-3 hypotheses and their associated frame-level probabilities for Dutch are of higher quality than in the case of Romanian.

% \subsubsection{Weak Supervision and Self-labeling}

% Table \ref{tab3} shows the impact of weak-supervision using video metadata versus self-labeling, both for an encoder-decoder model. For the low-resource setup in Dutch and Romanian respectively, while the weakly-supervised system improves the supervised-only baseline by XX\%, the self-labeled model far exceed it by a XX\% relative WER reduction over the baseline. When combined during the pre-training phase with a mixing ratio of 0.3, there is barely any noticeable difference to the final WER showing that the self-labeled system has already squeezed all the benefit from the diverse audio input without any need for further weak labels. 

For the encoder-decoder models, table \ref{tab3} compares weakly supervised learning using \texttt{data-ws} with meta-data targets and self-labeling using both \texttt{data-ws} and \texttt{data-large} audio. In both cases, a supervised \texttt{fine-tune} stage is used at the end of training. While weakly-supervised models improves over their respective baseline models from table \ref{tab1} by about 10\%, they don't come near observed gains from self-labeling. For clean and noisy conditions the smaller audio size of \texttt{data-ws} is enough to get all the gains from self-labeling. However, performance in extremely noisy conditions continues to improve with larger volume of unlabeled audio for both languages. Combining weak-supervision on \texttt{data-ws} and self-labeling on \texttt{data-large} barely makes any noticeable change compared to the self-labeling alone, showing that the model has achieved all its gains from diverse audio conditions without utilizing much of the information in the textual meta-data. This can be attributed to the meta-data quality and their weak relevance to audio content. 

\begin{table}
\centering
\setlength\tabcolsep{1.0pt}
 \begin{tabular}{cccccccc} 
\toprule
 \multirow{2}{*}{} &\multicolumn{3}{c}{Dutch}&& \multicolumn{3}{c}{Romanian} \\ 
 \cline{2-4}\cline{6-8}
 & clean & noisy & extreme && clean & noisy & extreme \\
 \midrule
% \multicolumn{8}{c}{Distribution matching}\\
% \midrule
Weak-supervision (WS) & 24.5 & 24.6 & 35.9 && 21.6 & 25.9 & 42.5 \\
%  \midrule
\midrule
% \multicolumn{8}{c}{Sequence matching}\\
% \midrule
SL (\texttt{data-ws}) & 18.4 & 18.8 & 28.7 && 13.3 & 15.6 & 28.2 \\
%  \shortstack{SL\\with WS audio} & 18.4 & 18.8 & 28.7 && 13.3 & 15.6 & 28.2 \\ 
SL (\texttt{data-large}) & 18.4 & 18.5 & 27.9 && 13.1 & 15.6 & 27.3 \\
%  \shortstack{SL\\full set} & 18.4 & 18.5 & 27.9 && 13.1 & 15.6 & 27.3 \\ 
\midrule
0.7 SL (\texttt{data-large}) \\+ 0.3 WS & 18.5 & 18.6 & 28.1 && 13.1 & 15.4 & 28.0 \\ 
% \shortstack{combined\\0.7 SL + 0.3 WS} & 18.5 & 18.6 & 28.1 && 13.1 & 15.4 & 28.0 \\ 
\bottomrule
\end{tabular}
%18.41	18.59	28.44	29.39	avg epochs 20-39	18.67%	28.54
\caption{WERs of weakly-supervised (WS) and self-labeled (SL) enc-dec models using both \texttt{data-ws} and \texttt{data-large} as well as their combination.}
\label{tab3}
\vspace{-3mm}
\end{table}

%Table \ref{} also shows the impact of increasing the size of the unlabeled video data used during self-labeling. %{\color{red} add comment}. %All numbers in table \ref{} are for encoder-decoder models. 
% \subsubsection{Iterative Self-labeling}

% Inspired by recent work of iterative distillation of student models with progressively more advanced students \cite{noisystudent}, we demonstrate the impact of applying this recipe to hybrid frame-level distillation in Table \ref{tab4}. Each iteration accounts for a round of knowledge distillation and supervised LFMMI. In the 0-iteration case, the models are only trained on supervised data and increasing parameters does not improve WER. To maximize the use of unlabeled data, we repeat the distillation process while continuously increasing the size of the student and see WER improvements with every iteration up to a final 16\% relative reduction. Conventional knowledge distillation uses a teacher larger than the student, but we show here that training a larger student can also work. This is potentially due to the extra parameters being better able to leverage the large amounts of unlabeled data in a  training setup with data augmentation methods like SpecAugment and dropout  as proposed in \cite{noisystudent}.

Motivated by the success of early iterative self-labeling \cite{wessel_05} and more recently in the computer vision tasks \cite{noisystudent}, table \ref{tab4} demonstrates iteratively distilling progressively larger student models that are distilled as teachers in following steps. Each iteration accounts for a round of frame-level distillation of a hybrid model followed by LFMMI fine-tuning using supervised data. In 0-iteration cases, baseline models of different sizes are trained on the data-augmented supervised data with almost no difference in WER. Each iteration of self-labeling added about 3\% to 4\% reduction in WER compared to a single iteration, pushing the overall gain from 10\% to 16\% on average compared to the baseline supervised system. Larger student models have greater capacity to benefit from the diverse audio conditions in the \texttt{data-large} set.

\begin{table}
\centering
\setlength\tabcolsep{2.0pt}
 \begin{tabular}{ccccccc} 
\toprule
 \multirow{2}{*}{} &\multicolumn{3}{c}{Dutch} \\ 
 \cline{2-4}
  &clean & noisy & extreme  \\
 \midrule

 0-iteration 80M parameters & 23.6 & 23.3 & 32.8 \\
 \midrule

 0-iteration 150M parameters & 23.2 & 23.1 & 33 \\
 \midrule

 0-iteration 240M parameters & 23.5 & 23.2 & 32.9 \\
 \midrule
 1-iteration 80M parameters & 21.3 & 20.7 & 29.8 \\
\midrule
 2-iteration 150M parameters & 20.5 & 20.0 & 28.6 \\ 
\midrule
3-iteration 240M parameters & 19.7 & 19.4 & 27.6  \\ 
\bottomrule
\end{tabular}
\caption{WERs of the Dutch test sets using Top-3 iterative frame-level distillation for up to three iterations of relabeling \texttt{data-large}.}
\label{tab4}
\vspace{-3mm}
\end{table}

\section{Discussion and Related Work} \label{related-work}
Reducing the amount of manual transcription for building speech recognition systems has been a constant research theme over the past three decades \cite{Zavaliagkos_98, Kemp1999, csl01_limsi, wessel_05, JHU_2012, hank2013} given the revolutionary increase and access to unlabeled data and computational resources. Starting as far as the mid-1990s, early research work used a bootstrap model trained on few hours of supervised data to generate labels for a larger set of audio data, with confidence filters applied to remove wrong transcriptions \cite{Zavaliagkos_98, Kemp1999, Bhuvana_05, ma_bbn_06}. 
TV shows and news closed captions were one of the early forms of audio metadata used either to bias the LM during decoding or as direct labels \cite{lafferty_96, csl01_limsi, chan_icassp2004, wang_ICASSP07}. Visual grounding used as weak-labels for learning audio representations in\cite{glass_16, ttic_may2017}.
Starting from as little as 1.2 hours of transcribed data, \cite{wessel_05} used iterative re-labeling and self-labeling while progressively increasing the model size to halve the overall system WER.
Motivated by the work on model compression \cite{caruana_06} and distillation \cite{ba_14, hinton_15}, and using 1 million of unlabeled personal assistant data, \cite{hari_2019} showed relative WER reduction of about 10\% to 20\% compared to supervised baseline. 
Incorporating the LM information within a sequence discriminative loss function for self-labeling was proven to be a difficult task given the noise in labels \cite{wang_ICASSP07, vimal_15, vimal_18}. On the other hand, seq2seq models showed greater grace working with noisy labels both for personal assistant domains \cite{tara_sslearning} and audio books \cite{kahn2019self, wei_awni_19}. With the full 960h Librispeech \cite{librispeech} data for building a strong initial model, \cite{gab_19} showed that using the larger Libri-light \cite{librilight} unlabeled data in the same domain in a self-labeling seq2seq setup significantly improves over the best supervised system achieving a new state-of-the-art for Librispeech.

%To benchmark progress in self- and semi-supervised learning, Libri-light \cite{librilight} offers a larger set of unlabeled audio data that is in the same domain as the 960h large scale Librispeech \cite{librispeech} benchmark, with three near-zero training sets of 10h, 1h, and 10min showing where self-labeling is seriously challenged. With the full 960h Librispeech data for building a strong initial model, \cite{gab_19} showed that using the full Libri-light unlabeled data in a self-labeling seq2seq setup significantly improves over the best supervised system achieving a new state-of-the-art for Librispeech.

\section{Conclusion and Future work}
\label{conclusion}
%On 27,000 hours and 58,000 hours of unlabeled Dutch and Romanian public social media video data respectively, self-labeled encoder-decoder speech recognition models with sequence-level distillation showed WER reduction of more than 20\% relative to the strongest data-augmented supervised baseline. These gains are the biggest compared to other frame-level self-labeling and weak-supervision using video metadata as distant labels. Moving beyond our current minimum of 150 hours of initial supervision for self-labeling teachers, future work will focus on methods that preserve the same level of WER while using one or two orders of magnitude lower volumes of initial labeled data.

In this paper, we investigated self- and weakly-supervised training with hybrid, encoder-decoder and CTC-based models for low-resource speech recognition on Dutch and Romanian public social media videos with about 300 and 150 hours of supervised data respectively.  
Using 27,000 hours and 58,000 hours of unlabeled Dutch and Romanian data, self-labeled encoder-decoder speech recognition models with sequence-level distillation achieved WER reduction of more than 20\% relative to the strongest data-augmented baseline. 
%These gains are the biggest compared to frame-level and sequence-level distillation on hybrid and CTC models. 
While weakly-supervised trained models using video meta-data as distant labels brought 10\% relative improvement over the encoder-decoder baseline, they didn't come close to the self-labeling gains. Combining the distant and self-labels barely made any difference, showing that the combined system didn't utilize the contextual metadata information.
We additionally observed that on hybrid models, frame-level distillation using Top-3 frame-level labels provided the same level of gains as the sequence-level distillation which utilizes a language model during label generation. While the encoder-decoder models seem to benefit from the implicit sub-word language model of the  self-labels, further research is needed to realize these benefits on other systems.
Moving beyond our current minimum of 150 hours of supervised data, future work will focus on methods that preserve the same level of WER while using one or two orders of magnitude lower volumes of supervised data. 
%This scenario is also more likely to present a greater importance on language model and contextual metadata.
% We believe the encoder-decoder model learns a longer range language model on the sub-word units of the self-labels and therefore provides the largest gains of 20\% on average compared the best baseline supervised system.
% Confirming its sensitivity to transcription quality \cite{vimal_18, hari_2019}, using self-labels for LFMMI training, even with the large amount of audio data, is not as useful as using the much smaller supervised data in the discriminative training stage of hybrid models.
% Interestingly, using Top-3 frame-level labels for distillation provided the same level of gains as the hybrid sequence-level distillation which utilizes a language model during label generation.
% Combining weak-supervision on \texttt{data-ws} and self-labeling on \texttt{data-large} barely makes any noticeable change compared to the self-labeling alone, showing that the model has achieved all its gains from diverse audio conditions without utilizing much of the information in the textual meta-stream. 
% Each iteration of self-labeling added about 3\% to 4\% reduction in WER compared to a single iteration, pushing the overall gain from 10\% to 16\% on average compared to the baseline supervised system.

\bibliographystyle{IEEEtran}
\bibliography{refs.bib}
\end{document}